\documentclass[letterpaper,12pt]{article}

\usepackage{verbatim,url}
\usepackage{amsmath,amssymb}
\usepackage{graphicx}
\usepackage{graphbox}
\usepackage{subfigure}
\usepackage{tabularx}
\usepackage{color,balance,url}
\usepackage{authblk}
\usepackage[numbers]{natbib}

\newcount\Comments 
\definecolor{darkgreen}{rgb}{0,0.6,0}

\author[1]{Stav Yanovsky}
\author[2] {Nicholas Hoernle}
\author[1]{Omer Lev}
\author[2,1]{Kobi Gal}

\affil[1]{Ben-Gurion University, Israel}
\affil[2]{University of Edinburgh, UK}

\begin{document}
\title {One Size Does Not Fit All: A Study of Badge Behavior in Stack Overflow}

\date{}
\maketitle

\begin{abstract}
Badges are endemic to online interaction sites, from Question and Answer (Q\&A) websites to ride sharing, as systems for rewarding participants for their contributions. This paper studies how badge design affects people's contributions and behavior over time. Past work has shown that badges ``steer'' people's behavior toward substantially increasing the amount of contributions before obtaining the badge, and immediately decreasing their contributions thereafter, returning to their baseline contribution levels. In contrast, we find that the steering effect depends on the type of user, as modeled by the rate and intensity of the user's contributions. We use these measures to distinguish between different groups of user activity, including users who are not affected by the badge system despite being significant contributors to the site. We provide a predictive model of how users change their activity group over the course of their lifetime in the system. We demonstrate our approach empirically in three different Q\&A sites on Stack Exchange with hundreds of thousands of users, for two types of activities (editing and voting on posts). 
\end{abstract}


\section{Introduction}
Many online platforms rely on the motivation of volunteers rather than on paid workers to create content~\citep{ipeirotis2014quizz}. Examples include Wikipedia, Reddit, Question and Answer (Q\&A) sites like Stack Overflow, and citizen science platforms in which non-experts collaborate with scientists to accelerate scientific discoveries~\citep{simpson2014zooniverse}. Moreover, social-media websites also rely, to a large degree, on users for creating content. 

Keeping users productive and motivated is essential to the success of such peer production sites~\citep{simpson2014zooniverse}. One of the most commonly used incentive mechanisms used by these sites are badge systems, which provide users with credentials that display skills and achievements on the site~\citep{seaborn2015gamification,Cavusoglu}. Badge systems partition the set of participants into ``status classes'' that reflect their contributions according to a particular metric~\citep{immorlicaSS15}. When administered successfully, badge systems can influence users' behavior and direct them towards types of activities encouraged by the system designers~\citep{anderson2014engaging}. The use of badges can also signal expertise or experience to users as well as to wider communities~\citep{hickey2015badges}.

Despite the massive use of badges in online communities\footnote{e.g., \url{http://duolingo.wikia.com/wiki/Achievements}}, Q\&A sites\footnote{e.g., \url{https://askubuntu.com/help/badges}}, ridesharing\footnote{e.g., \url{https://blog.lyft.com/badge-glossary/}} and more, our understanding of the interplay between user behavior and badge design is still lacking. 

Much previous work has focused on badges' ``steering'' effect~\citep{anderson2013steering,li2012quantifying}. That is, 
users' contribution levels rise as they get closer to the threshold that is required for obtaining the badge, and experience a sharp decline following it, returning to their baseline contribution levels.

In this paper we show that the steering effect is not homogenous, but varies across different types of users.
Our data is taken from the \emph{Stack Exchange} platform, which hosts a collection of Q\&A websites, each devoted to a different topic and includes hundreds of thousands of users. We focused on the largest project of the platform, the programming-related \emph{Stack Overflow} (we also analysed data from several of the smaller projects, like Ask Ubuntu and TeX-LaTeX but as results are similar we will often refer mainly to Stack Overflow). 

We examine two common and general tasks on Stack Overflow: voting on other people's posts, and editing others' posts for corrections and clarifications.   
We chose these badge types because they represent  fundamental and popular activities in Stack Exchange, hence our insights are representative of the general 
		population of users. We show that the user population can be clustered into 3 separate groups, differentiated by the frequency and intensity of their work on the task. We show that users in each of these different groups experience steering in a different way, and we examine their short and long term reaction to receiving badges. In particular, we find that some users do not return to their baseline levels of contributions after receiving the badge. For these users, badges act as a \emph{catalyst} for long-term activity on the platform, creating a sustained level of activity over an extended period of time. 

We provide a computational model for predicting whether a user will decrease her level of contribution to a lower activity group on the site following a badge award. This model can inform the design of personalized intervention methods to increase the contributions of such users. Our work has insights for system designers in showing that badges are not a ``one size fits all'' incentive and it suggests ways to adapt existing badge designs to the diversity of user behavior. 

This paper extends a prior conference submission which focused on the SO project in several ways. First, we extend our results to an additional class of voting actions and show that our model generalizes well to this new class. Second, we examine the interaction between different types of badges. Lastly, we extend the related work section to provide more comprehensive background of badge design and user behavior in the literature.

\section{Related Work}
Badge design and the effects of badges on people's interactions with online systems has been studied in the social and computational sciences. \citet{hickey2015badges} outlined key guidelines for successful badge design, such as transparency (the badge system should be known and understood by all users, badges should be visible), interactions (badge systems are more successful in settings where there is a high degree of interaction between participants), and uniqueness (badge systems should be the sole incentive mechanism in the domain setting). Different sites use different badge designs, each with its own particular purpose~\citep{easley2016incentives}. For example, some badges award users for contributing valuable content while others award users for being among the most prolific contributors of the site.

%
%

A few studies model the influence of badges on user behavior in social media and Q\&A sites~\citep{zhang2016badge,Cavusoglu,anderson2013steering,li2012quantifying,halavais}. 
Most central to our work are the studies by \citet{anderson2013steering} and \citet{li2012quantifying} which describe the ``steering'' phenomena towards a badge boundary: As users approach the threshold of the required number of actions needed to earn a badge (day zero), they increase their contributions needed for the badge.
The total amount of actions that are influencing the earning of the badge increases significantly in the days that are prior to earning the badge, in a comparison to earlier days and the days after getting the badge. \citet{bosu} analyzed the social network inferred from participants' questions and answers on the site, inferring how to obtain reputation scores quickly on the cite. Other works have studied the role of personality traits, as inferred by their SO messages, on their badge behavior~\citep{Papoutoglou,bazelli2013personality}. \citet{bornfeld2017gamifying} provided a longitudinal analysis of feedback (votes and responses)   as a mechanism for increasing contributions among newcomers in five Stack Exchange communities, including SO.

\citet{anderson2013steering} present a mathematical model which describes the deviation of distribution over user actions before and after receiving the badge. 
They use the model to demonstrate the steering effect of badges on user voting behavior on Stack Overflow, and a few empirical studies followed~\citep{GB13}.
However, our results paint a more complicated picture than these, as we show that the steering effect is not homogeneous, and differs across different types of users. We also   study the long-term effect of badges over the lifetime of interaction of users in the system. 




Several works have studied badges in the context of academic courses. \citet{anderson2014engaging} studied badge design and its effect on student behavior in a large student forum in a massive open online course (MOOC). They showed that placing several badges of smaller value that are well dispersed in the course can be more effective than having a single badge of higher value.

\citet{hakulinen2015effect} showed that rewarding students taking a computer science course with achievement badges motivated students and 
encouraged desired study practices.
\citet{charleer2013improving} studied different visualizations of badges that reward students' forum activity in a course.
They 
compared personal dashboards, where students can observe each other's badge achievements, and an augmented version in which students could discuss the badge achievements with each other. They showed that the personal dashboard improved awareness of the course's goals, while the interactive visualization improved the students' collaboration and reflection on the coursework. 

\citet{abramovich2013badges} used an intelligent tutoring system that notified students whenever they earned a badge and explained the reason for earning it. 
This approach led to an improvement in students' engagement and a decrease in counter-productive behavior, when compared 
to badge-less tutoring systems. 

Badges have been used in gamified apps, as systems designers use game design elements to 
improve user engagement and experience \citep{seaborn2015gamification,deterding2011gamification,hamari2015gamification}.
Gamification is studied from the aspects of psychology~\citep{linehan2015gamification}, game theory~\citep{easley2016incentives,jain2009designing}, and economics~\citep{hamari2015gamification}.
Common gamification elements includes the use of points, levels, leaderboards, time constraints, badges and more \citep{seaborn2015gamification}. \citet{jia2016personality}
present a survey study investigating the relationships among individuals' personality traits and perceived preferences for various gamification elements.

Badge design has also been studied from a game theoretic 
perspective~\citep{easley2016incentives}. \citet{immorlicaSS15} studied badge design mechanisms aiming to maximize the total contributions made to a website. Users exert an effort (which carries a cost) to contribute and, in return, are rewarded with badges. Badge valuations are determined by the number of users who earn each badge. 
\citet{aoyagi2010information} considered the role of badge design as what feedback to provide to two agents in a two-round contest. The agents 
can expend some amount of effort in each round, with a noisy mapping between effort and score. 
Both papers characterize the equilibrium strategies that need to hold in their respective model. 

Finally, several works criticized the use 
of badges as an incentive mechanism~\citep{deci2001extrinsic}, claiming that badges may undermine intrinsic motivation. In particular, \citet{kobren2015getting} found that students tend to drop out of e-learning systems just after obtaining the necessary amount of questions to achieve the relevant badge.  

\section{Setting and Research Questions}
The setting for this work is the Stack Exchange (SE) platform, a network of 173 question-and-answer (Q\&A) websites on topics in diverse fields, in which users post and respond to questions\footnote{\url{https://stackexchange.com/}. User data from SE is freely available.}.

For the investigation that follows, we chose the following three projects in SE which vary widely in their topics and in the number of active users.

\begin{description}
\item [Stack Overflow] (about 9,000,000 users) deals exclusively with programming and is the biggest and most popular site on SE; 
\item [Ask Ubuntu] (about 474,000 users) a site for users and developers of the Ubuntu operating system;
\item [TeX-LaTeX] (about 67,000 users) a site for users of \TeX, \LaTeX, ConTeXt, and related typesetting systems. 
\end{description}

\begin{figure}
\centering
\includegraphics[width=\columnwidth]{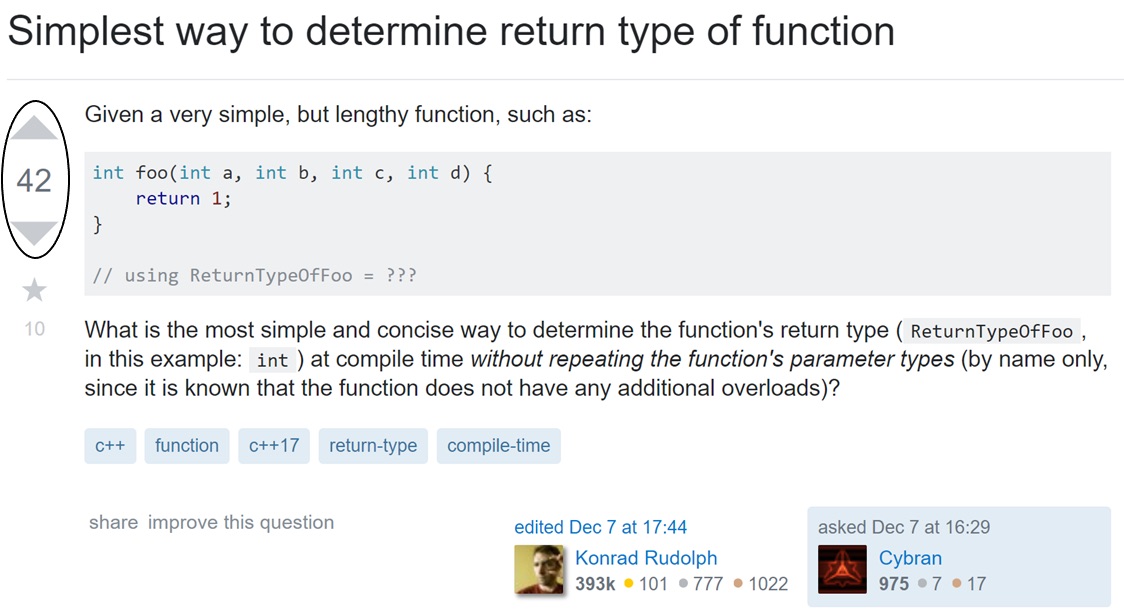}
\includegraphics[width=\columnwidth]{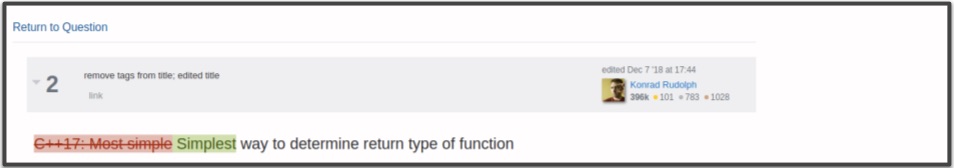}
\caption{An example of a post in the Stack Overflow project (top) which has been up-voted 42 times (left) and an edit activity to the post (bottom)}
\label{fig:au}
\end{figure}

The primary purpose of each SE site is to enable users to post questions so other users can answer them. In order to improve both questions and answers, users can also edit and comment on each other's posts. This allows users to correct existing posts (e.g., in the case of typos) or to provide insights about the post content. Users can also vote (up or down) for posts, providing a  reputation score that is displayed on the user profile. By performing actions such as posting questions or answers, voting on posts and editing existing posts, users can increase their reputation score on the site, and unlock various privileges~\citep{bosu}. Figure~\ref{fig:au} shows an example of a post (top) which has been upvoted (left) with an edit action (bottom) in SO. Of course, the multitude of users (as in other crowdsourced projects, such as Wikipedia), are casual ones, who only access content on the website but do not actively contribute to it. 

All SE projects employ badges to incentivize contributions by users. 
There are more than 100 different badge types in SE, each divided to three ranks in increasing order of importance: bronze, silver and gold. To differentiate between them, SE has aliases for the different type of badges that one can earn. For example, edit-type badges in SE sites use the aliases ``Editor'' for a bronze badge value, ``Strunk \& White'' for a silver badge value, and ``Copy Editor'' for a gold badge value. The badges rewarding voting actions are ``Supporter'' or ``Critic'' for one's first vote (depending on if it was an up-voter or down-vote) as the bronze badge; ``Civic Duty'', given after voting 300 or more times (regardless of the vote type) is the silver badge; and ``Electorate'', given to users after voting on 600 questions (of which at least 25\% are on questions), is the gold badge. 

\subsection{The Edit Badges Data}

\begin{figure}
	\centering
	\includegraphics[width=5cm]{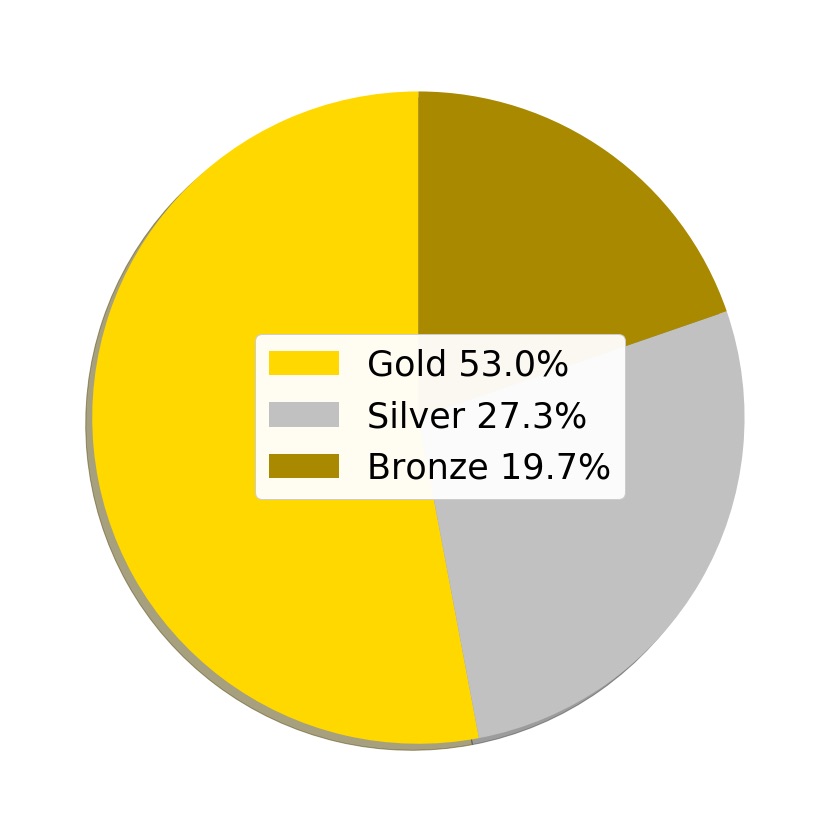} 
	\caption{Percentage of edit-actions contributions for winners of bronze, silver and gold badges}
	\label{fig:pieChart}
	\end{figure}
	
We will initially focus on the edit badge. Our analysis is based on data of user interactions on the SO project from September 2008 and up to May 2019. Figure~\ref{fig:pieChart} shows the percentage of edit-action contributions made by winners of the different badge types. 

We are particularly interested in the 14,276 users who achieved the silver badge; 2,687 of these users went on to achieve the gold badge. Together, this group made the vast majority of edit contributions to the site. Understanding how these users behave can inform the design of future incentive mechanisms in the site. Although winners of the bronze badge make up the vast majority of the user population (more than 90\%), they do not provide a substantial contribution of edit actions to the site. In Section \ref{sec:discussion} we explore how the badge system might be designed to encourage more participation from the users who only achieved the bronze badge.

\subsection{Research Questions}
Our research focuses on the way badges affect different users. We are particularly interested in the \emph{steering} phenomenon identified 
by \citet{anderson2013steering}, where users tend to increase their contribution rates as they approach the badge goal. 
We are also interested in modeling how badges affect people's long term behavior on the platform. 

We study the following research questions:
\begin{enumerate}

\item Is steering one-size-fits-all? Do different user populations experience steering in different ways? 
\item How do badges affect the long term behavior of individual users throughout the lifetime of their interaction in the system?
\item Does the steering effect identified by \citet{anderson2013steering}, and our findings on the edit badge, extend to other types of actions and SE projects, beyond voting actions in Stack Overflow (SO)?

\end{enumerate}

We first focus, for the first two questions, on analyzing badge behavior for edit actions. Frequent edit type actions in SE include correcting grammatical errors or misspellings in a post, or adding explanations to the existing post content. The thresholds for achieving the bronze, silver and gold badges are a single edit action, 80 edit actions, and 500 edit actions respectively. These threshold values are standardized across all of the SE projects. Our hypothesis was that different types of users ``steer'' differently, i.e., they vary in the extent to which they respond to badges. Moreover, we believe that individual users vary in how the badges affect their behavior throughout the course of their lifetime on the system. For the third question, we widen our angle to more SE projects and other badges.

\section{Question 1: Is Steering One-Size-Fits-All?}\label{sec:q1}
In this section we study whether steering effects differ between different types of users, as exhibited by their behavior on the site. Intuitively, users with similar numbers of contributions may still exhibit widely different activity styles. For example, consider two users; one of them performs 5 edit actions each day of the week and the other performs 35 edit actions on Sunday nights. In total, both users contribute an equal number of edit actions per week, but clearly, they exhibit different behavior patterns on the site.

To distinguish between such users, we define two measures: 

\begin{description}
\item [Work Consistency:] The median number of days spent editing in a week.
\item [Work Intensity:] The median number of edits that a user makes in a day, given the user makes at least one edit.
\end{description}

We chose these two measures because (1) 
they provide a general description of user activity in SE that does not depend on the action type itself (e.g., the number of characters changed or added in an edit activity); (2) they provide a simple and succinct way to differentiate between user behavior in the site. 
For example, a user who was active 
for three days in the first week of activity, five days in the second week of activity, and three days in the 
third and final week of activity will have a consistency value of 3. Similarly, a user who produced two edit actions in the first day of activity, ten edits in the second day of activity, and five edits days in the third and final day of activity will have an intensity value of~5. We considered the median number of edit actions rather than the mean because the distribution over edit actions per day is right-skewed and is highly affected by outliers.

\subsection{Inferring User Groups}
\begin{figure}
\centering
\includegraphics[width=\columnwidth]{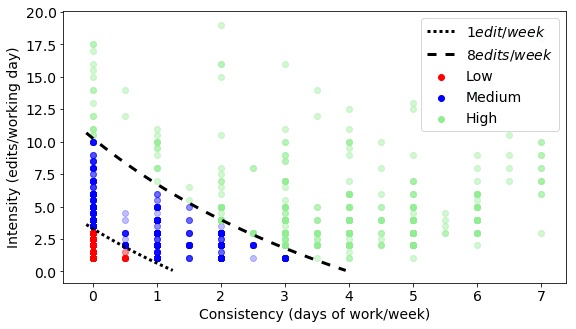}
\caption{Scatter plot of user activity showing three user groups revealed by k-means (K=3). Groups are distinguished using colors and 
boundary curves.}
\label{fig:kmeans_3}
\end{figure}

Using the notions of work consistency and intensity, we wish to group users into distinct clusters of activity. In order to do so, we utilize the k-means algorithm. Figure~\ref{fig:kmeans_3} plots the work consistency and intensity for the gold and silver users in SE (recall this group contributes the vast majority of edits in the system). The algorithm used the two measures to cluster users into three groups of activity: low, medium and high. Groups are distinguished in the figure using colors and boundary curves.

The default distance metric used in k-means is euclidean distance between datapoints and cluster centers. However this does not constitute a good metric for the purposes of differentiating between groups with different levels of contributions. For example, when using euclidean distance, if we consider a cluster with (consistency, intensity) centroid $(4,5)$, and two users with (consistency, intensity) measures of $(1,5)$ and $(7,5)$, the users would exhibit the same distance from this cluster center. However they display different activity levels. The first user works one day a week and would complete an expected total of 5 edits per week, while the second works every day in the week and would complete an expected total of 35 edits per week. 

To this end, we define a custom distance that captures the expected number of posts per week directly. For two users $a$ and $b$ and their respective intensity and consistency ($I_a, C_a, I_b$ and $C_b$), the distance $d$ is defined as:
\begin{equation}
d(a, b) = ABS(I_{a} \times C_{a}-I_{b}\times C_{b})
\end{equation}\label{eq:distance_metric}
\noindent The group centers can thus also be interpreted in terms of common number of edits per week.

The clusters are formed in a transformed parameter space using the following steps. First, we drop the users with consistency and intensity scores greater than the 99.9 percentile in each case. This corresponds to 15 users who all had an intensity greater than 20. Second, we normalize the data by dividing by the maximum value and adding 1 to offset the effect of 0 values.

We choose $k=3$ to facilitate the interpretation of the clusters. We aim to describe general trends in the data while still accounting for the fact that users are interacting with the system in unique ways. Increasing the cluster parameter $k$ to $4$ simply had the effect of splitting the high-activity group into two, thus complicating the further analysis unnecessarily.

Using the modified distance metric, the k-means algorithm reveals the following three types of user groups. 
The low activity group describes ``dabbler'' users whose activity is characterized by low consistency and intensity levels (contribute generally less than
a single edit per week). The medium activity group describes users who exhibited a medium level of intensity and consistency (contribute generally between one and 8 edits per week and rarely work more than 4 days for any given week). The high activity group describes ``busy bee'' users who exhibited a high level of intensity (contribute generally more than 8 edits per week and regularly work on more than 3 days in any given week). Returning to our example from above with the two users described by (consistency, intensity), we can see that the first user is in the medium activity group and the second user is in the ``busy bee'' group.

Table~\ref{tbl:silver_gold_activity_groups} shows the number of silver and gold users in each activity group. 
As shown in the table, low-activity users make up the vast majority of the user population, followed by the medium and high activity user groups. Gold users make up just 16\% of low activity users, 44\% of the medium user group, and over a half of the high user group. Thus, despite the lower rate of contributions
exhibited by the low/medium activity groups, they still make up a substantial part of the contribution. 

\begin{table}
	\centering
	\begin{tabular}{lccc}
		\hline
		& \textbf{Low} & \textbf{Medium} & \textbf{High} \\ \hline
		\textbf{Silver users} & 10,022 & 1,380 & 287 \\
		\textbf{Gold users} & 1,198 & 1,119 & 370 \\ \hline
		\textbf{Total} & 11,220 & 2,499 & 657 \\
		\hline
	\end{tabular}
	\caption{Number of gold and silver users in each activity group}
	\label{tbl:silver_gold_activity_groups}
\end{table}

\begin{figure}
\centering
\includegraphics[width=0.79\columnwidth]{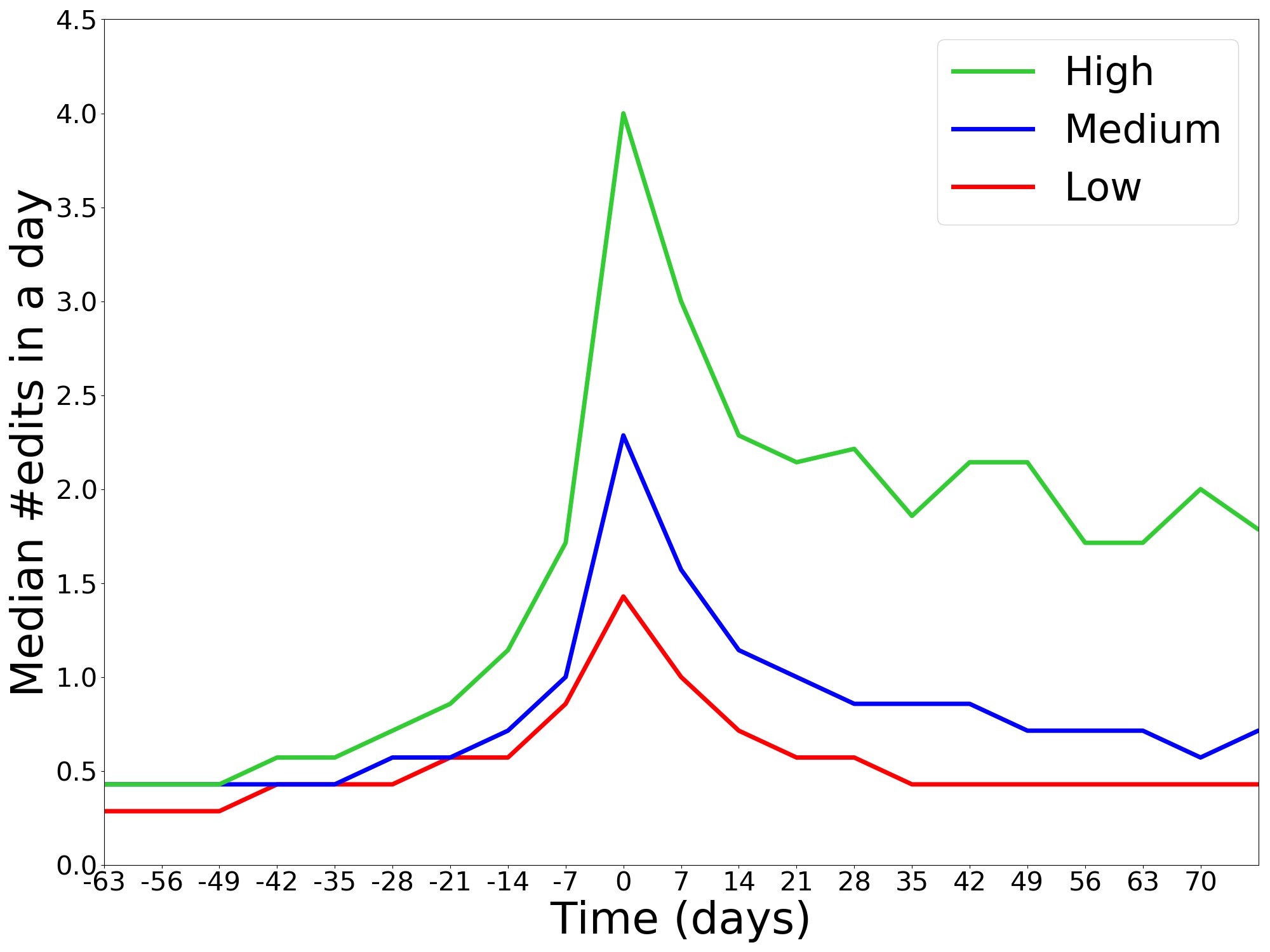}
\caption{Median number of edits per day, centered around the day-zero for achieving the silver badge}
\label{fig:steering_effect_on_intensity}
\end{figure}

\subsection{Separating the Badge Effect}
Figure~\ref{fig:steering_effect_on_intensity} plots the 
contributions of the different engagement groups over time, relative to day zero, when the silver badge was awarded to the user. As shown in the figure, several days 
before day zero, there is no discernible difference in the contribution levels of the three groups, as their usage pattern seems to be identical across the clusters. 

However, in the few days just before day zero, high-activity users experience a sharp rise in the number of edits per day leading into the silver badge and maintain this high rate of editing for a number of weeks (!) after the receiving the badge. The behavior of these users runs counter to \citet{anderson2013steering}'s prediction that after receiving a badge, users will return to their default levels of activity. In contrast, the other groups are far less affected by the badge design. Both low and medium-activity groups exhibited a smaller jump in contributions prior to day-zero (with low-activity also smaller than medium-activity), and a steeper decline after this day. However, medium-activity users did settle on contribution levels slightly above their previous, pre-badge, default level, while low-activity users returned to their previous work habits. Notice that the low activity group is the largest and it therefore dominates the trends when all of these users are aggregated. It is only when we analyze these groups individually that this nuanced behavior becomes apparent. 

At the peak of the contribution level, there is a highly statistically significant difference between the three levels of contributions; $p \ll 1\times 10^{-4}$ one-way ANOVA. Before obtaining the silver badge the difference between the levels of contributions is not statistically significant (30 days before day zero - $p = 0.206$ one-way ANOVA), while after obtaining it, the difference stays significant (30 days after day zero - $p \ll 1\times 10^{-4}$ one-way ANOVA).

\section{Question 2: How do Long Term Group Dynamics Change in the Presence of Badges?}

In this section we explore how users change their behavior over time. To this end we track whether and how low, medium, and high-activity users change group types before and after receiving badges in the system. Thus, we divide users into groups for different parts of their life cycle in the system (before badge/after badge, etc.).

\begin{figure}
	\centering
	\includegraphics[width=0.9\columnwidth]{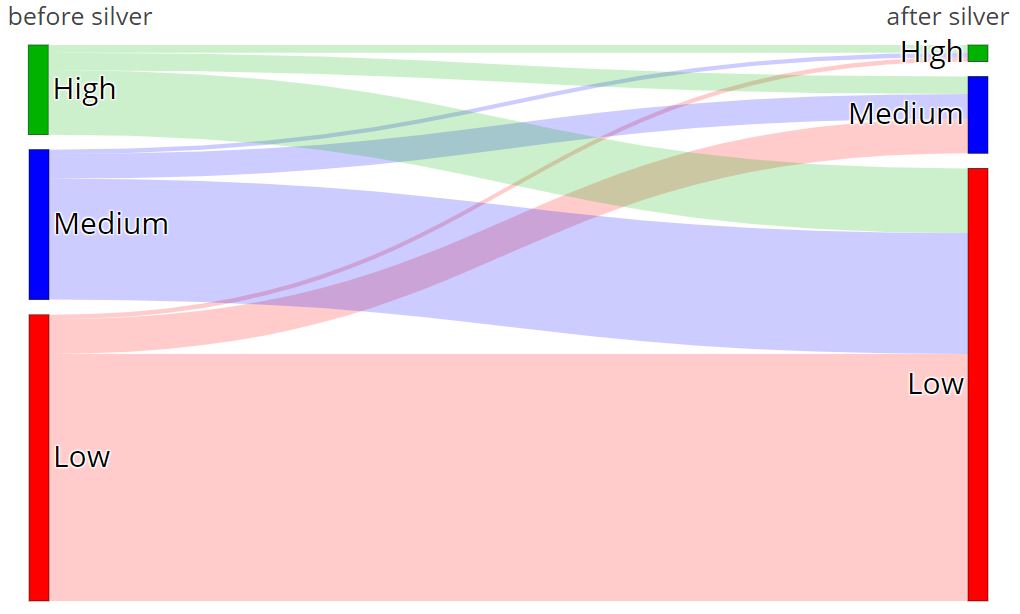}
	\caption{Flow between user groups before (left) and after (right) getting the silver badge}
	\label{fig:silver_shift}
\end{figure}

We begin by tracking the long term behavior of users who received a silver badge but not the gold badge (these users are responsible for 27\% of user contribution 
on the site, see Figure~\ref{fig:pieChart}). Figure~\ref{fig:silver_shift} is a Sankey diagram tracking the flow of these 
users between the different group types. As can be seen, the vast majority of users (including medium and high activity users) became, once the badge was awarded, low activity users. Only a tiny minority of low and medium users became high-activity users. The behavior of these users agrees with the theory of steering, in that they returned to their usual work patterns following the silver badge acquisition. In the discussion section we discuss the implications of this 
behavior to facilitating badge design.

\begin{figure}
	\centering
	\includegraphics[width=0.9\columnwidth]{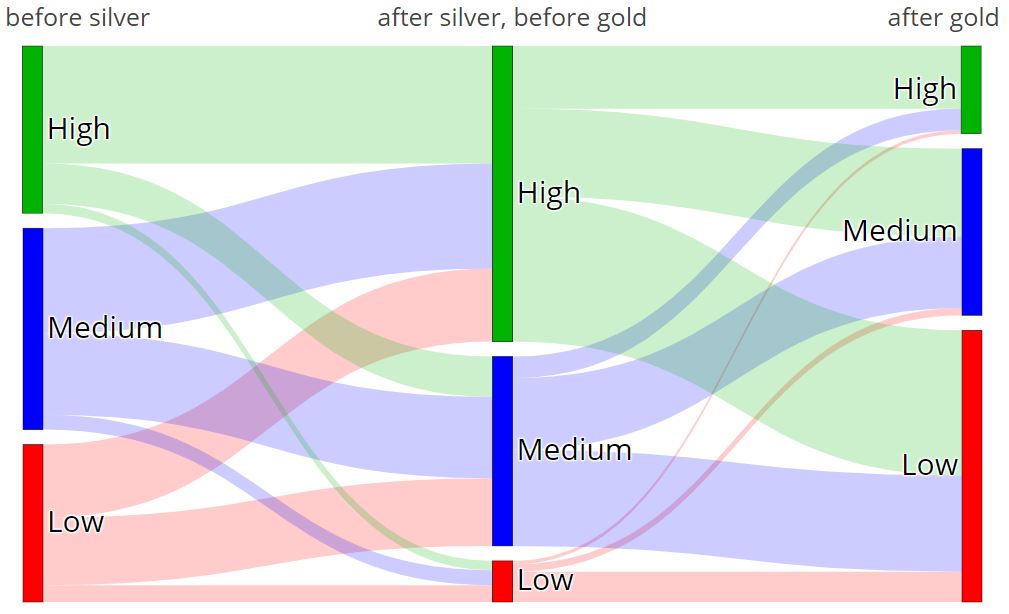}
	\caption{Flow between user groups before getting the silver badge (left), after getting the silver and before getting the gold badge (middle) and after getting the gold badge (right).}
	\label{fig:gold_shift}
\end{figure}

We now turn to track the long term behavior of users who received the gold badge (these users are responsible for 53\% of user contribution on the site). Figure~\ref{fig:gold_shift} shows the flow between groups for these users before obtaining the silver badge (left), after obtaining the silver and before obtaining the gold badge (middle) and after getting the gold badge (right). As shown by the figure, the user shifting between groups is quite different than that of users who did not obtain the gold badge (Figure~\ref{fig:silver_shift}). The users depicted here did not return to their normal routine once they achieved a silver badge. Instead, they generally increased their activity -- an overwhelming majority either stayed at the same activity level or increased it, and for the medium and low-activity users, a sizable majority strictly increased their activity levels to become high-activity users. However, once the gold badge was achieved, most of the users reverted to the same steering behavior identified 	by \citet{anderson2013steering}, and their activity levels decreased significantly.

\subsection{Are Some Badges Gateways to Other Badge?}
We now explore a different long term dynamic, on how two badges interact with one another. One badge is the edit badge, which we have explored so far, and the other, the voting badge, which we will delve into in section~\ref{votingBadgeAnalysis}. We use the available data on the period between September 2008 to May 2019 in SO.

\begin{table}
	\centering
	\begin{tabular}{cc|ccc|}
		\cline{3-5}
		& & \multicolumn{3}{c|}{\textbf{Edit Badges}} \\ \cline{3-5} 
		& & \textbf{no badge} & \textbf{silver-edit} & \textbf{gold-edit} \\ \hline
		\multicolumn{1}{|c|}{{\textbf{Vote Badges}}} & \textbf{no badge} & & 1992 & 96 \\
		\multicolumn{1}{|c|}{} & \textbf{silver-vote} & 65822 & 5228 & 391 \\
		\multicolumn{1}{|c|}{} & \textbf{gold-vote} & 11091 & 5934 & 2676 \\ \hline
	\end{tabular}
	\caption{Amounts of users obtaining at least on of the four discussed badges in SO}
	\label{tbl:votes_edits_so}
\end{table}

Table~\ref{tbl:votes_edits_so} summarizes the amount of users in SO who obtained at least one of the following badges: edit-silver, edit-gold, vote-silver or vote-gold. Not surprisingly, the number of users with vote badges is significantly larger than the amount of users who earned edit badges. This is, presumably, since voting badges are far easier to achieve: while a single edit requires entering a separate screen and making a change to the existing text, a vote requires only a single click. However, what is evident from the data is that achieving an edit badge is usually correlated with also achieving a vote badge. Indeed, 87\% of the users who obtained the silver-edit badge also achieved the silver-vote badge. On the other hand, only 15\% of the users who obtained the silver-vote badge also achieved the silver-edit badge. 

\begin{figure}
	\centering
	\includegraphics[width=\columnwidth]{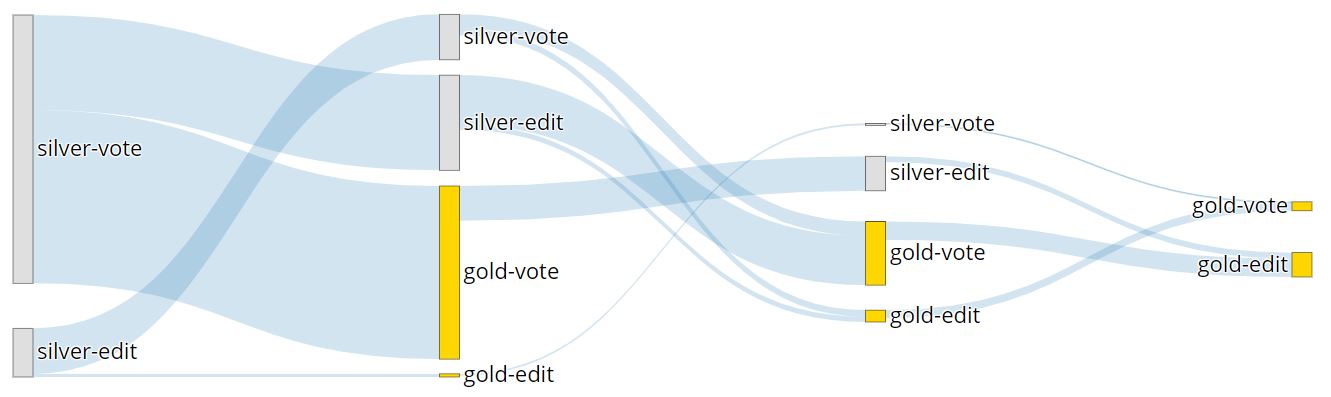}
	\centering
	\caption{Chronological order of obtaining the edit and vote badges. People who received only a single badge (silver-vote or silver-edit) are not shown.}
	\label{fig:vote_edit_order}
\end{figure}

When looking at the chronological order of obtaining these badges (Figure~\ref{fig:vote_edit_order}), some patterns emerge. The vast majority of the users begin by collecting the vote-silver badge. Out of the users who start from getting the silver-edit badge first, only a slight amount of users go on immediately for the gold-edit badge. Instead, most of them go towards achieving the silver-vote badge. Perhaps the gold-edit badge seems too hard to achieve at that point, so they choose a different badge as their next goal. 

Focusing on the most productive and engaged population (in terms of editing and voting), this population includes 2,676 users who earned gold badges for both voting and editing action types. These users performed about 14.5 million votes out of all 138 million vote actions that were done on SO (about 10\%). Amazingly, these users performed about 3,700,000 edits out of all 8,500,000 edit actions that were made on SO (about 43\%!). As shown by Figure~\ref{fig:vote_edit_order}, the most popular path for achieving the four badges is in the following order:(1) silver-vote, (2) silver-edit, (3) gold-vote and (4) gold-edit. These users first achieve all of the silver badges before moving on to achieving all of the gold badges. In addition, they almost always choose to achieve vote badges first.

Indeed, it seems these users are pursuing relatively short-term goals: choosing the action type that belongs to the most reachable badge in terms of its demands. According to this ``algorithm", the users begin with the easiest action type (voting) and the closest badge (silver, by default). At this point, the users prefer to pursue the silver-edit badge rather than or the gold-vote badge, since the silver badge is more accessible for them. Moreover, it seems that the relatively easy to achieve silver-vote badge serves as a gateway to further badge achievements.

\section{Question 2, Continued: Can We Model User Behavior?}

\begin{table}[t]
	\centering
	\small
	\begin{tabular}{lccccc}
	\hline
	\textbf{Features} & \textbf{\# Features} & \textbf{Accuracy} & \textbf{F1 Weighted} & \textbf{ROC AUC} & \textbf{Confusion Matrix} \\ \hline 
	\textbf{User (U)} & 10 & 0.704 & 0.693 & 0.640 & \begin{tabular}[c]{@{}c@{}}{[}7987 1549{]}\\ {[}2700 2140{]}\end{tabular} \\ 
	\textbf{Edits (E)} & 9 & 0.680 & 0.633 & 0.569 & \begin{tabular}[c]{@{}c@{}}{[}8680 \textbf{856}{]}\\ {[}3739 1101{]}\end{tabular} \\ 
	\textbf{Temporal (T)} & 40 & 0.835 & 0.839 & \textbf{0.875} & \begin{tabular}[c]{@{}c@{}}{[}7174 2362{]}\\ {[}\textbf{12 } 4828{]}\end{tabular} \\ 
	\textbf{U +E + T } & 59 & \textbf{0.837} & \textbf{0.842} & \textbf{0.875} & \begin{tabular}[c]{@{}c@{}}{[}7236 2300{]}\\ {[}42 4798{]}\end{tabular} \\ \hline
	\end{tabular}
	\caption{Prediction results for user decreasing contribution levels after day zero, using different combinations of features.}
	\label{tbl:predictions}
\end{table}

Our machine learning model seeks to 
understand which users are susceptible to decreasing their work habits after obtaining the (silver) badge, so that we can target these users and tailor incentive solutions for them in order to increase their motivation (see Discussion Section). Therefore, it is critical to know in advance which users will decrease their work on the platform and which users will be maintaining the same levels of work. 

\subsection{Feature Extraction for Prediction Task}
A user is represented by a vector that includes three distinct families of predictor variables: \textbf{user features}, \textbf{edit features} and \textbf{temporal features}.

\begin{description}
\item[User Features] include features specific to the user such as her age, length of activity history in the system as well as the number of other SO badge achievements won prior to the date when the silver badge was awarded. 
\item[Edit Features] included features that summarize the user's edit history. These features include the ratio of edit actions to other actions performed on the system, statistics about what part of the posts they edit (e.g. title or content) and how long are the comments describing each edit. 
\item[Temporal Features] included features summarizing the user's consistency and intensity measures from different periods of time from the user's interaction history. We represent the history as a vector of mean consistency values for each week of the user's lifetime in the system, up to day zero (and similarly for intensity values). To measure changes in these two metrics, we average the consistency and intensity through time for 3, 5 and 10 weeks prior to achieving the badge and the 3, 5, and 10 weeks of a user's activity in the system . For example, for a given consistency history $(c_1,\ldots,c_n)$ of $n$ weeks of activity, these features average the consistency values 
$(c_1, \ldots c_3)$ for the first 3 weeks of activity, and similarly for the first 5 and 10 weeks of activity prior to the badge.
We also create features for 3, 5 and 10 weeks of the user's interaction history prior to receiving the badge. These features average the consistency values $(c_{n-2},\ldots, c_n)$ for the last 3 weeks of activity before day zero, and similarly for the last 5 and 10 weeks of activity. 
Similar features are defined relating to the intensity of users as well. This allows the prediction to harness relative changes in the user's behavior at different points in time relative to day zero. Another important feature in this family of features is the user's activity group prior for obtaining the badge.
\end{description}
The prediction task is whether the user will decrease her contributions and move to a lower group type after receiving the silver badge. Specifically, we predict whether high activity users descend to the medium activity group, and whether medium activity users descend to the low activity group.

\subsection{Prediction Results}

We used the XGBoost classifier algorithm~\citep{chen2016xgboost} for this prediction task, and tried different combination of the following parameters: the number of used trees, the maximum depth of the trees and the learning rate of the algorithm. We used ten fold cross validation, with standard deviation of 
results between runs smaller than 0.01 for all measures. 
As can be seen in Table~\ref{tbl:predictions}, in most regards, using all feature types produced the best results. However, note that most of the prediction quality comes from using the temporal based features. The user and edit features have relatively weak prediction ability (a combination of them showed a negligible increase in prediction ability), and using the temporal variables alone seems to give excellent results without requiring any extensive knowledge of the users themselves or their particular editing habits.

Using the user and edit features alone led to a rather small number of errors in one direction -- fewer people were mistakenly predicted to decrease their activity, when they did not (false positive). However, using the temporal features alone, while increasing false positives, almost eliminated the error of predicting people will not decrease their activity when the did (false negative). When trying to prevent people from decreasing their activity, false negatives are more important to focus on, because presumably, many engaged users will brush off attempts to engage them further, while users who are not targeted to prevent their dropping-out, are forever lost to the system.

\section{Question 3: Does Steering Generalize?}

\begin{figure}
	\centering
	\includegraphics[width=0.7\columnwidth]{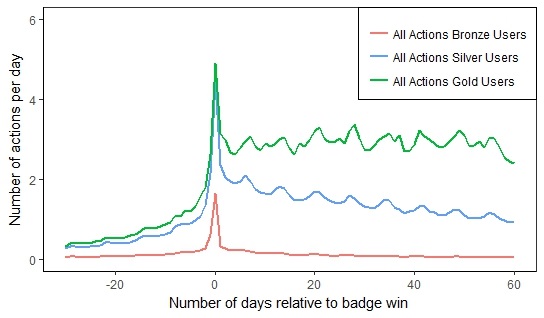}
	\includegraphics[width=0.7\columnwidth]{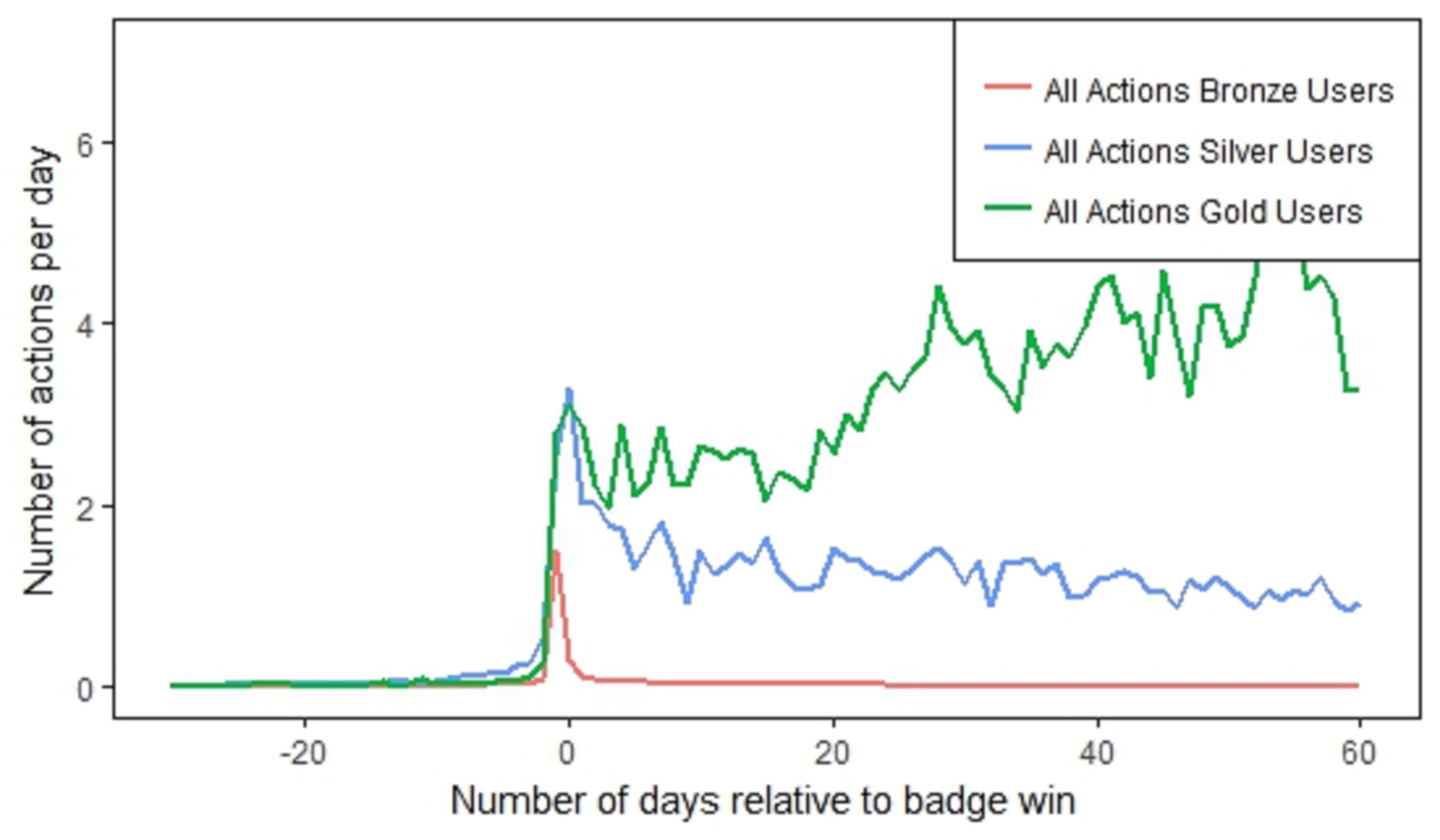}
		\includegraphics[width=0.7\columnwidth]{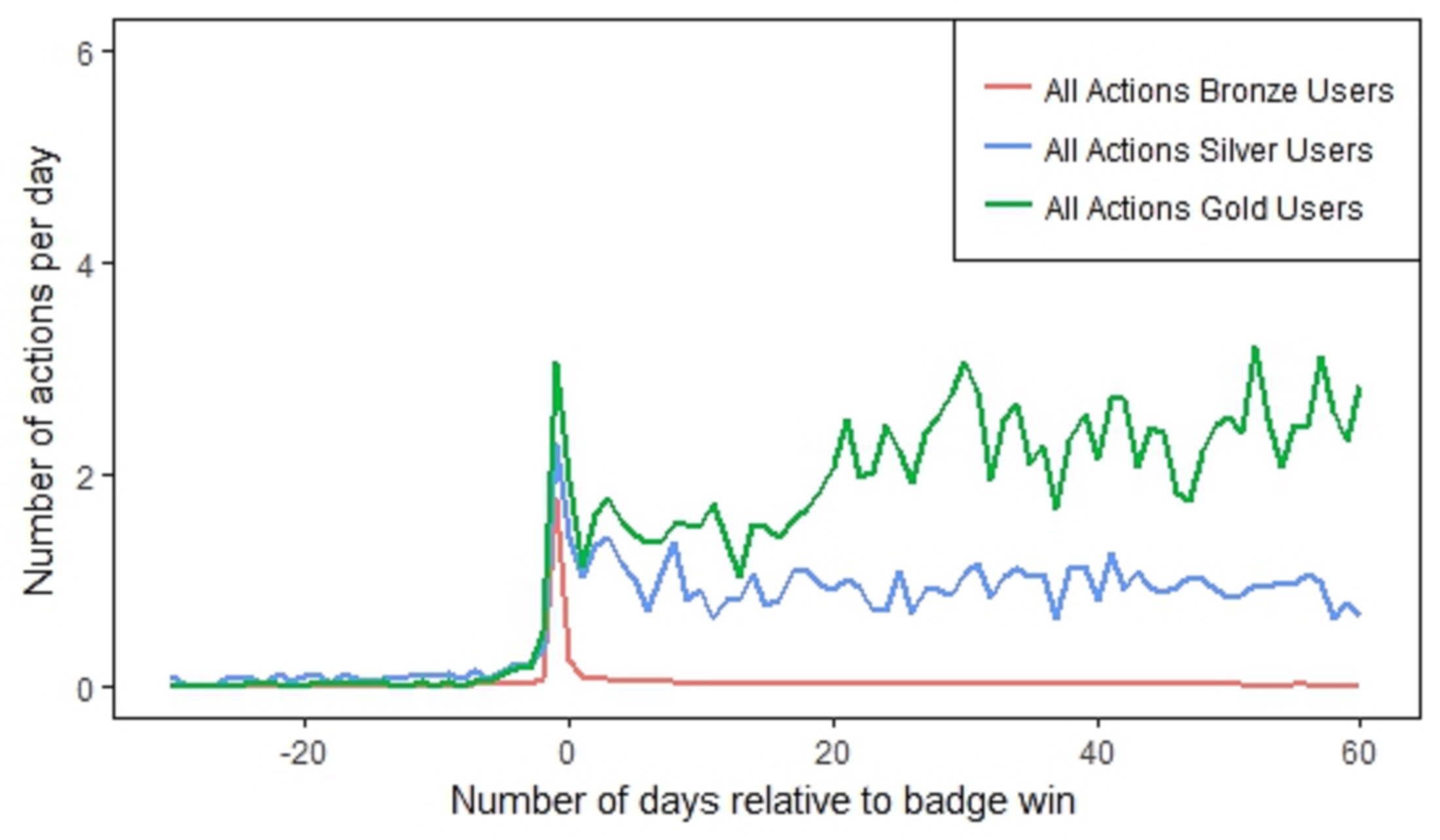}
	\caption{Average number of edit contributions as a function of distance from day-zero for obtaining silver badge on the Stack-Overflow (top), AskUbuntu (middle) and \TeX-\LaTeX~SE sites (bottom).}
	\label{fig:steering generlizes}
\end{figure}
In this section we study whether steering generalizes to other Stack Exchange projects and other badges. 

\subsection{Generalizing to Other Stack Exchange Projects}
Figure~\ref{fig:steering generlizes} shows the average number of edit contributions as a function of the number of days from day-zero (the day in which the silver badge was obtained by the user) for all three projects: Stack-Overflow (top), Ask Ubuntu (middle) and TeX-LaTeX (bottom) SE sites. The negative numbers to the left of day-zero give the time in days prior to obtaining the badge. Accordingly, the region to the right of the axis, with positive numbers, is the time after getting the badge. 

As shown by the figure, in all projects, users exhibit a sharp rise in activity as they approach day zero, and following this day they exhibit a steep decline in contribution, returning to their default rates of activity. The figure confirms that the steering effect identified by \citet{anderson2013steering} holds in other projects as well. Moreover, the steering effect was not limited to only voting actions and badges but appears to hold for editing action and badges as well.

Furthermore, as analyzed in Section~\ref{sec:q1}, in these projects as well, the behavior of each badge group is different, and there is no ``one size fits all'' phenomena. The machine learning model which we have built for SO works in these projects as well, and with similar performance.

\subsection{Generalizing to the Vote Action}\label{votingBadgeAnalysis}

As mentioned earlier, the two measures we defined in order to characterize user activity in the SE domain are work consistency and work intensity. We can generalize these measures from dealing with edit actions to deal with other action types, including vote actions. This is the same activity  which was analyzed by ~\cite{anderson2014engaging}. We want to show how the analysis that dealt with the edit actions behavior can be generalize to vote actions, and point out the differences between behavior under both of these badge types. The data regarding the vote actions in each day is confidential, and was achieved using a collaboration with the Stack Overflow academic research department. The data is from the beginning of 2017, and we used all users who obtained the silver-vote badge after April 2017, so that we will have enough data for the time period of before obtaining the badge. 

\begin{figure}[t]
	\centering
	\includegraphics[width=\columnwidth]{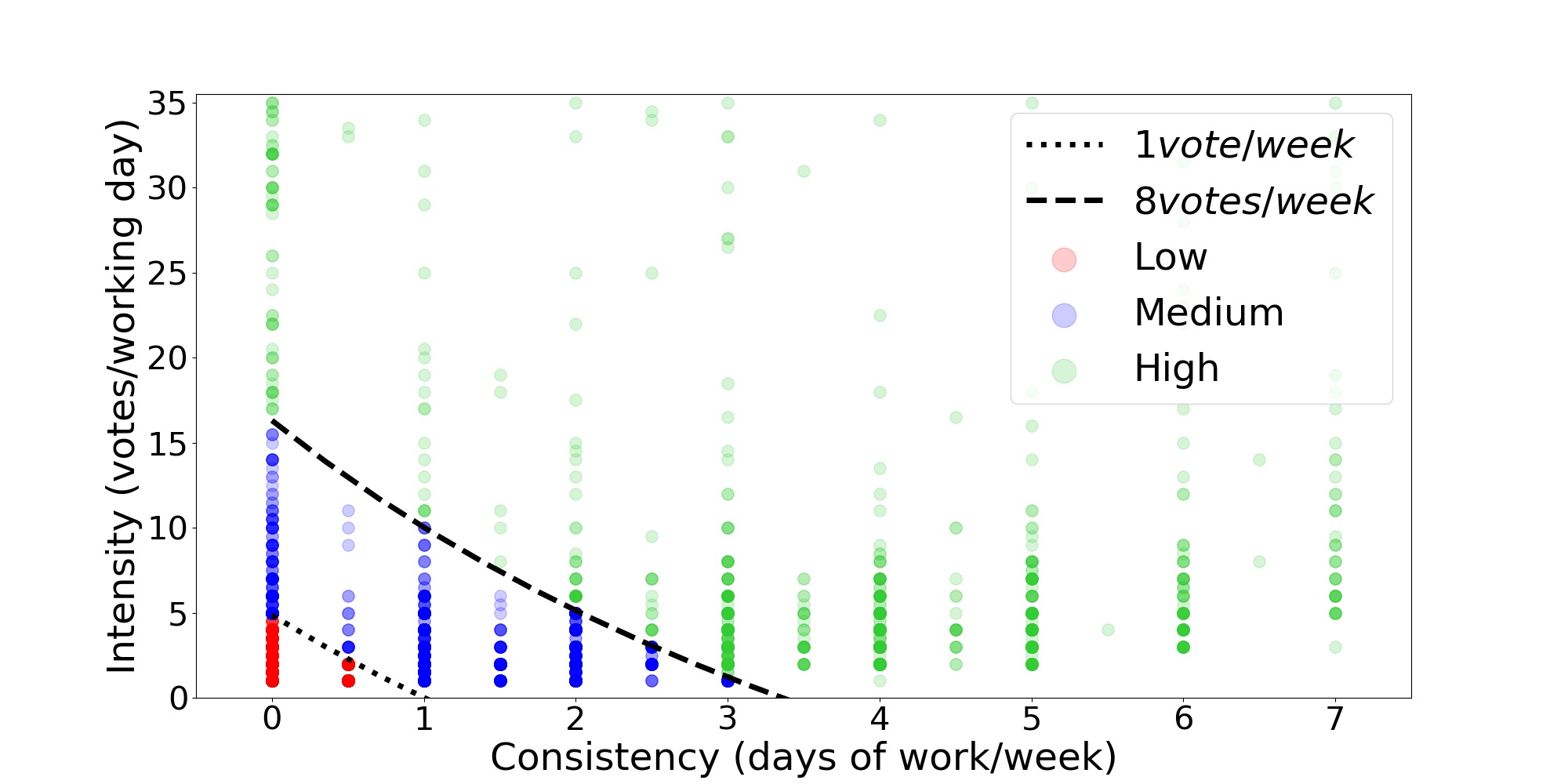}
	\caption{Scatter plot of user voting activity showing three user groups revealed by k-means (K=3). Groups are distinguished using colors and boundary curves}
	\label{fig:kmeans_3_votes}
\end{figure}

\begin{table}[t]
	\centering
	\begin{tabular}{lccc}
		\hline
		& \textbf{Low} & \textbf{Medium} & \textbf{High} \\ \hline
		\textbf{Silver users} & 9,049 & 12,476 & 992 \\
		\textbf{Gold users} & 368 & 993 & 960 \\ \hline
		\textbf{Total} & 9,417 & 13,469 & 1,952 \\
		\hline
	\end{tabular}
	\caption{Number of gold-vote and silver-vote users in each activity group}
	\label{tbl:silver_gold_activity_groups_votes}
\end{table}
We used the same consistency/intensity measures we used for the edit badges, and we use the k-means algorithm using the distance metric as described in~\ref{eq:distance_metric} to cluster all users who got at least the silver-vote badge into three groups of activity: low, medium and high. Figure~\ref{fig:kmeans_3_votes} shows the clusters that were created, and Table~\ref{tbl:silver_gold_activity_groups_votes} describes the amount of silver and gold users in each group of activity. The number of users in each cluster has changed dramatically from the edits analysis; here the largest group is of the medium activity and not the low as it was in the edit analysis. We can also see that the boundaries of the clusters ``shifted'', and now a user with a consistency median value of 0 and a intensity median value of 5 will be assigned to the low activity group, rather to the medium activity group in the edit analysis (same goes to the boundary between the medium and high activity groups). The reason for these changes is that voting is a much easier and more incidental action than editing. 

\begin{figure}[t]
	\centering
	\includegraphics[width=12cm]{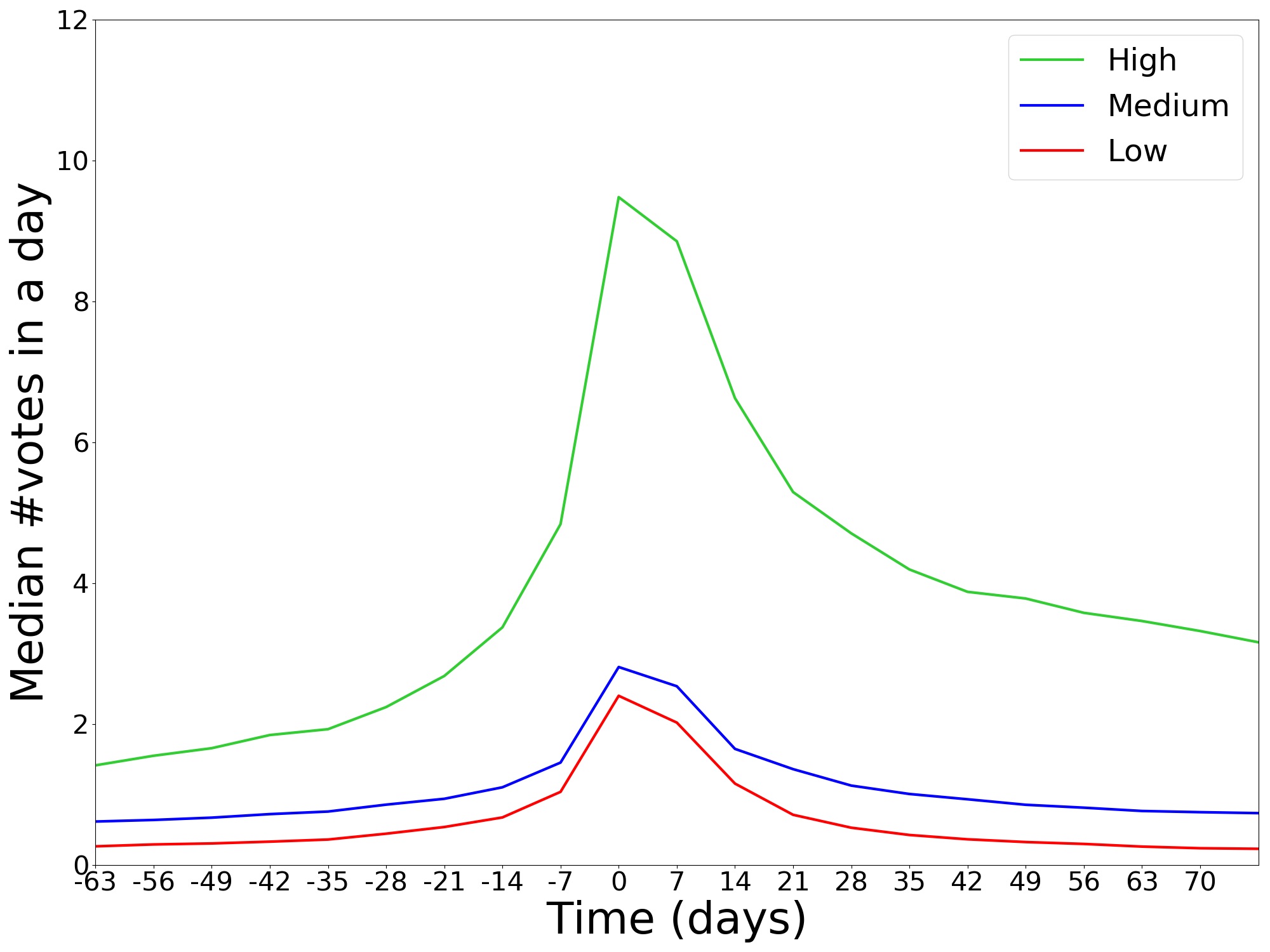}
	\caption{Median number of votes per day, centered around the day-zero for achieving the silver-vote badge}
	\label{fig:steering_effect_on_intensity_votes}
\end{figure}

As for the steering effect of the silver-vote badge, Figure~\ref{fig:steering_effect_on_intensity_votes} plots the contributions of the different engagements groups over time, relative to day-zero, when the silver-vote badge was awarded. We observe the same phenomena as in Figure~\ref{fig:steering_effect_on_intensity}, regarding to how different user populations steer when receiving a badge. The high activity users steers more than the other two groups, and they almost double the amount of votes in each day, comparing to their baseline activity prior to getting the badge. The other two groups of activity, medium and low, exhibit a far lower change in the activity on day-zero and clearly return to their baseline activity. Results regarding the voting data are also significant.

\begin{figure}[t]
	\centering
	\includegraphics[width=\columnwidth]{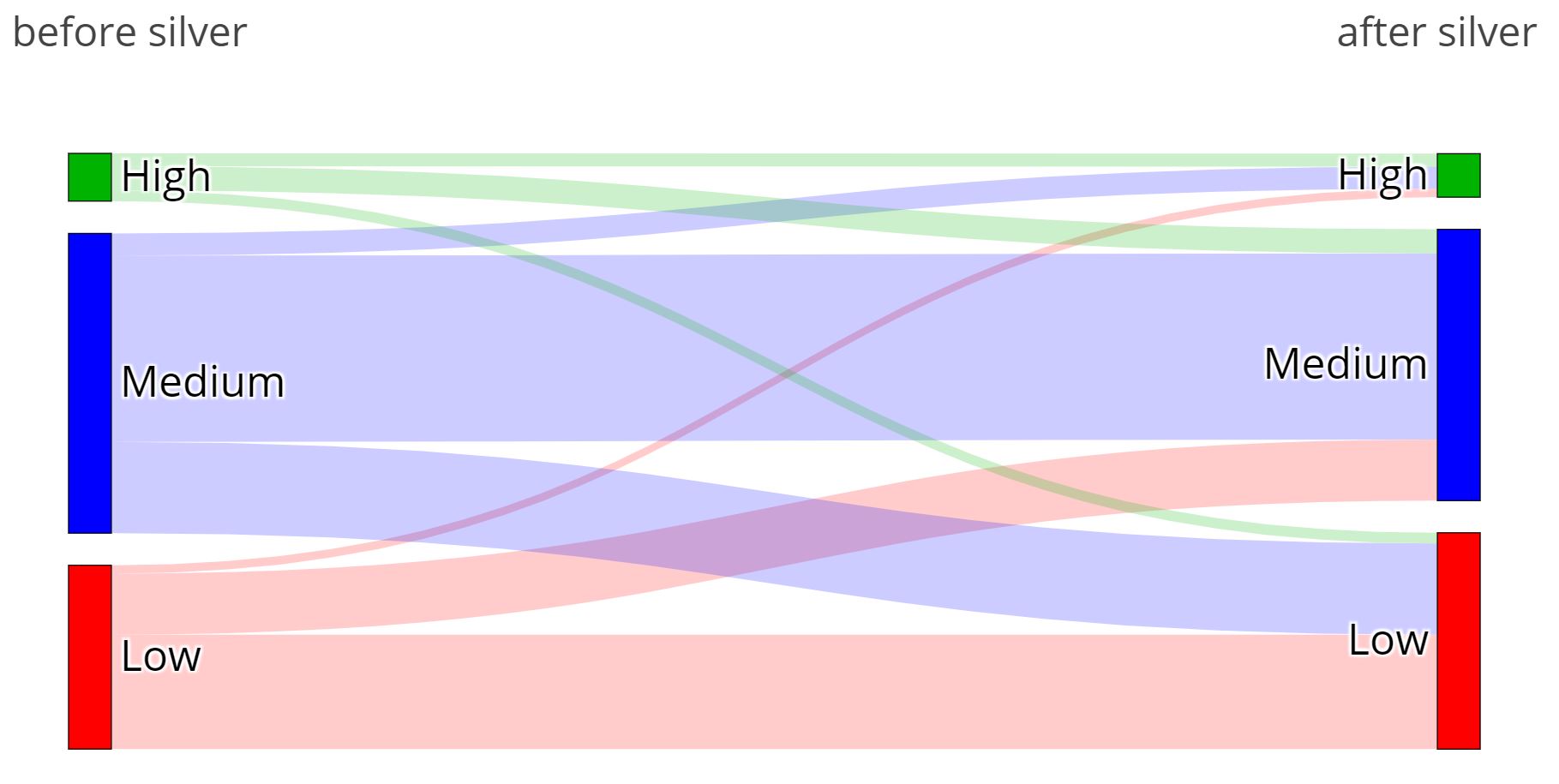}
	\caption{Flow between user groups before (left) and after (right) getting the silver-vote badge}
	\label{fig:silver_shift-votes}
\end{figure}

When looking at the users who received the silver-vote badge but not the gold-vote badge, we observe a different insight from the one we outlined for the same edit population. This group includes the users who made more that 300 votes in SO but less than 600 votes. In Figure~\ref{fig:silver_shift-votes}, we track the flow of these users before and after obtaining the silver-vote badge. Here, unlike we have seen in Figure~\ref{fig:silver_shift}, the amounts of users in each group do not seem to change dramatically. Getting the silver-vote badge does not make users vote less than they did before getting it. On the other hand, getting the silver-vote badge also does not make them vote more than they did before. We can say that these users are indifferent to receiving the silver-vote badge.

When looking at the users who were able to get the gold-vote badge, we observe the insight we found regarding the gold-edit achievers (see Figure~\ref{fig:gold_shift}). Here the vast majority of the gold-vote users belongs to the high activity group in the time period of between getting the silver-vote and gold-vote badges. Since the task of voting is easy for them, and they got some reward for doing it (the silver-vote badge), they highly increase their vote actions and become (or stay) high activity users. For them, the silver-vote badge had a positive effect, it made them act more. Again, when the gold-vote badge is achieved, we observe a decrease in voting activity; the amount of users in each activity group is close to as it was before getting the silver-vote badge.
\begin{figure}
	\centering
	\includegraphics[width=\columnwidth]{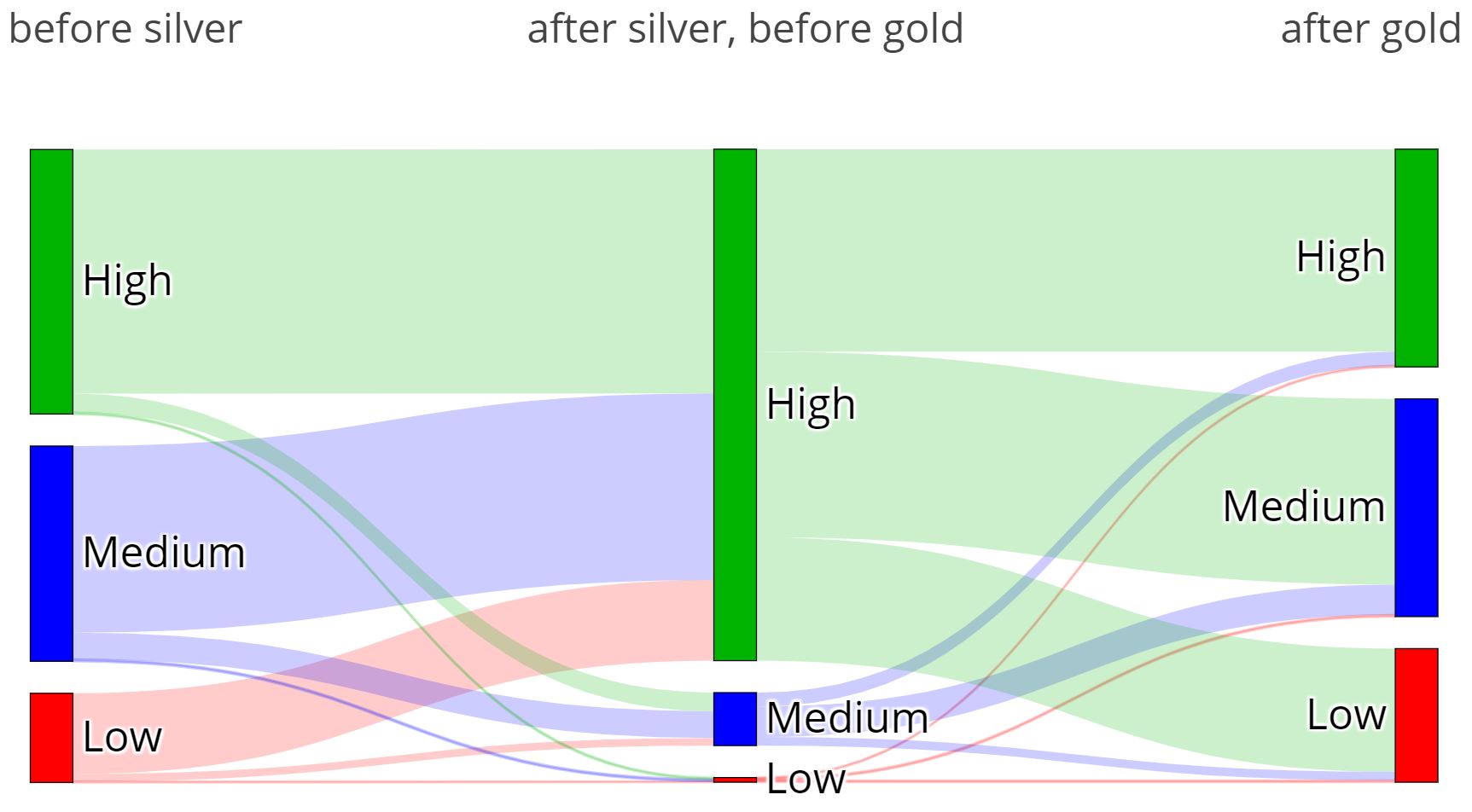}
	\caption{Flow between user groups before getting the silver-vote badge (left), after getting the silver-vote and before getting the gold-vote badge (middle) and after getting the gold-vote badge (right)}
	\label{fig:gold_shift-votes}
\end{figure}

\section{Discussion \& Future Work}
\label{sec:discussion}
Our main results in this paper elaborate and expand previous research on badges (and in particular,~\cite{anderson2013steering}):

\begin{enumerate}
\item Reaction to badges varies greatly between different user populations. In particular, large sections of users (e.g., our low-activity ones) register a very small reaction to badges at all, while others show a reaction that is at odds with model predictions, as they \emph{increase their work after receiving the badge}.

\item Engagement patterns of users can be an effective predictor, at least to some degree, of future badge reception (e.g., high-activity users and the gold badge). Classification of users based on their working habits may be beneficial for understanding which people might benefit from different incentives.
\end{enumerate}

We observe a far more complex interaction between badges and user behavior than noted previously. Our results indicate that while for many users, working intensely to receive a badge can be a one-time thing, for some users (who are the most productive ones, from the platform's point of view), badges have a different meaning. These users' behavior seems to indicate that once they receive their first meaningful badge, it encourages them to participate in the badge environment, and they want to achieve more badges, until there are no more to achieve. The badge seems to be the \emph{catalyst} for such a process, as prior to being awarded the silver badge, these users had much lower activity levels. However, even for these types of users, the badge system is meaningful, as once they have received a gold badge, they slowly drop off the system. 

Our observations here could be applied in different ways for different use-cases. When a high rate of participation is needed, it is clear that badges are failing to engage vast numbers of users, in particular the low-activity users, and more importantly -- those that do not even reach the stage of silver badges. Perhaps a different set of incentives might be needed for these users. On the other hand, badges are much more effective in motivating persistent behavior from a subset of users, and there is potential for badge behavior to focus on these users. For example, it may be beneficial to lower the threshold for awarding badges, to allow for this activity catalyst to reach users who may be ``sparked'' by it, but prior to receiving the badge, had such a low activity profile that they did not even reach the silver badge threshold. This would allow for an earlier identification of the other engagement groups, and hence to allow for focusing on the different incentives needed for each of these populations. Similarly, medium-activity users are increasing their activity after they receive the silver badge, taking quite a while to return to usual work patterns. Perhaps if the next badge was not so distant (500 edits for gold vs. only 80 for silver), they might have seen the badge goal as reachable, and become high-activity users, working to achieve it. 

One limitation of our approach   is that the empirical analysis does not distinguish between steered and non-steered populations in SO.  Indeed,  motivations that underlie 
	participants' activities in the site may be    influenced by  factors other than badge acquisition, 
	such as motivation to contribute to the community. The ``bump" in their activity can be explained by their natural  activity patterns and would also occur in the absence of steering.
	
\begin{figure}
	\begin{center}
		\includegraphics[width=10cm]{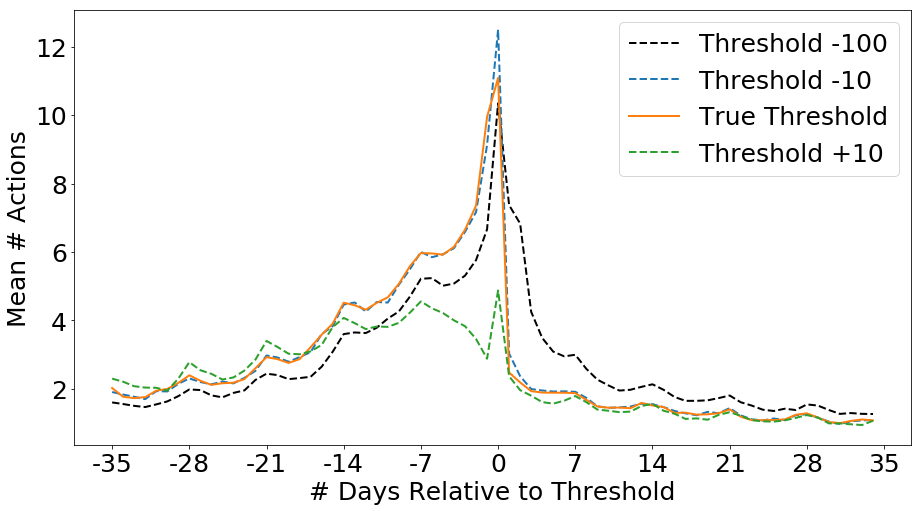}
	\end{center}
\label{fig:posthresh}
\caption{Mean activity as a function of time relative to threshold}
	\end{figure}

We answer this claim in two ways. First we  computed  activity graphs for  different acquisition thresholds for  one
	 of  the badges  in the study  the voting (``Electorate")
	badge. 
	The graph in  Figure~\ref{fig:posthresh} shows  the mean activity rate (count of vote actions per day) 
	of participants as a function of time  relative to  some 
	cumulative action threshold. The graph shows 
	activity curves that are centered on  the true Threshold (600 actions), 
	Threshold-10
	(590 voting actions); Threshold+10 (610 actions), and 
	Threshold-100 (500 actions).

The graph shows  there can be  \emph{significant differences} in the contribution patterns   depending on how the data is centered.  The  curve for Threshold+10  is   noticeably different from the other curves  with the  decline in contributions starting well before the     Threshold  cutoff point for this curve (the day in which 610 actions  were  completed). The spike of the Threshold+10 curve is much lower than the other curves. We argue that this is due to participants decreasing their activity rates once the true threshold is crossed,  
	which is in line with the steering phenomenon. 
	The curves for Threshold-10 and the true threshold are  similar in that they show an increase of activity leading into the threshold cutoff followed by a sharp decline in activity. The spike of these 
	curves is also similar, possibly because many people cross these two thresholds on the same day and get the badge. This trend is also in line with the steering 
	phenomenon. 
	The fact the   participation curves strongly depend on how  
	the data is centered with respect to the badge threshold  implies  that at least part of their  behavior can be attributed to the badge.

	Additional  support for  the effect of badges on   behavior was
	provided by  using the  approach by ~\citet{hoernle2020goalgradient} to 
	separate those participants 
	motivated by steering from those that do not deviate from their ``default" activity distribution.  Using this analysis, we discovered that about 38\% of medium-activity  	users in SO (at least 8 edits per week, rarely work more than 4 days per week)  are steered whereas only about 12\% of high-activity (above 8 edits per week, work more than 3 days per week) users are   steered.  On the one hand, this shows that high level contributors will not be affected by better badge design, but these users are already strong contributors to SO. On the other hand,    system designers can   target medium-activity  users, who are both   regular contributors to SO and respond positively to  badges.

The intricate interaction we uncover between badges and user behavior calls for much further research, and there is plenty left to do. For example, the role of multiple badges is not yet fully understood. We are extending our work to other types of badges (in particular, qualitative ones), and examine personalized badge structure. For qualitative badges, we may need to devise other measures for describing work progress towards a goal that go beyond consistency and intensity. We intend to experiment with different badge design schemes for engaging student learning in a soon-to-be-released MOOC. Our long term goal is to direct system designers on how to 
design a badge system optimally for a given platform.
Also, we are studying how to design intervention mechanisms that target individual users who are predicted to 
decrease their contribution level. To this end it is necessary to reason about the trade-off that is made between interrupting or frustrating an engaged user and between intervening with a user who might be disengaging with the platform~\cite{segal2018optimizing}.


\subsection*{Acknowledgments}
Thanks very much to Stack Overflow for making available the data on which this research was based. 
Nicholas Hoernle is supported by a commonwealth scholarship. 
Stav Yanovsky is supported by a grant from the Israeli Science Foundation number 773/16.

\bibliographystyle{plainnat}

\end{document}